\documentclass[prl,twocolumn,showpacs]{revtex4}

\usepackage{graphicx}

\begin{document}

\title{The fluid--fluid interface in a model colloid--polymer mixture:
Application of grand canonical Monte Carlo to asymmetric binary mixtures}

\author{R. L. C. Vink and J. Horbach}

\affiliation{Institut f\"{u}r Physik, Johannes Gutenberg-Universit\"{a}t,
D-55099 Mainz, Staudinger Weg 7, Germany}

\date{October 17, 2003}

\begin{abstract}
We present a Monte Carlo method to simulate asymmetric binary mixtures
in the grand canonical ensemble. The method is used to study the
colloid--polymer model of Asakura and Oosawa. We determine the
phase diagram of the fluid--fluid unmixing transition and the interfacial
tension, both at high polymer density and close to the critical point. We
also present density profiles in the two--phase region. The results are
compared to predictions of a recent density functional theory.
\end{abstract}



\maketitle

In colloid experiments, hard--sphere--like systems can be realized in
which the corresponding phase behavior is of purely entropic origin. An
example is the fluid--fluid unmixing transition that is observed in
solutions of colloids and non--adsorbing
polymers~\cite{asakura1954a,vrij1976a}. This transition is due to a
depletion effect~\cite{asakura1954a,vrij1976a,vrij1997a,roth2000a,aarts2002a}. Each
colloidal particle is surrounded by a depletion zone from which polymers
are excluded. When two colloids are close together, their depletion zones
may overlap, thereby increasing the free volume available to the polymers.
This results in an anisotropic pressure exerted by the polymers onto the
colloids which gives rise to an effective attraction between the colloids.
Such depletion forces also occur in mixtures of large and small hard
spheres, but in this case it is still debated whether phase separation
occurs~\cite{bhs}.

A simple model for colloid--polymer mixtures was first introduced
by Asakura and Oosawa~\cite{asakura1954a} and later independently
by Vrij~\cite{vrij1976a}. In this model (the so--called AO model)
colloids and polymers are treated as spheres with respective radii
$R_{\rm c}$ and $R_{\rm p}$.  Hard sphere interactions are assumed
between colloid--colloid (cc) and colloid--polymer (cp) pairs, while
polymer--polymer (pp) pairs can interpenetrate freely. This yields the
following pair potentials for the AO model:
\begin{eqnarray}
\label{eq:ao}
u_{\rm cc}(r) &=& \left\{
    \begin{array}{ll}
    \infty & \mbox{for $r<2R_{\rm c}$} \\
    0      & \mbox{otherwise,}
    \end{array}
  \right. \quad \nonumber \\
u_{\rm cp}(r) & = &
  \left\{
    \begin{array}{ll}
    \infty & \mbox{for $r<R_{\rm c}+R_{\rm p}$} \\
    0      & \mbox{otherwise,}
    \end{array}
  \right. \\
u_{\rm pp}(r) &=& 0 \nonumber,
\end{eqnarray}
where $r$ is the distance between two particles. The polymers thus
represent ideal polymer coils with a radius of gyration $R_{\rm p}$ which
can be realized experimentally in a $\theta$ solvent.

The AO model has been the subject of many studies in the framework of
density functional theories
(DFT)~\cite{brader2000,schmidt2002a,brader2002a}. In particular, these
studies yielded the phase diagram for a wide range of colloid to polymer
size ratios. Moreover, in the case of the fluid--fluid unmixing
transition, they predicted interfacial tensions that are roughly one
thousand times lower than those for simple liquids, in agreement with
experiments~\cite{hoog1999a,chen2000}. A drawback of the latter DFTs is
that they are mean field theories and thus cannot recover the 3D Ising
critical behavior observed, for instance, in a recent experiment on a real
colloid--polymer mixture~\cite{chen2000}.

Monte Carlo simulations are also well suited to study the phase behavior
of colloid--polymer mixtures. Recent simulations in the Gibbs ensemble
were performed to determine phase diagrams of the AO
model~\cite{dijkstra1997a} and also of a model that considers non--zero
interactions between the polymers~\cite{bolhuis2002a}. In the latter study
even quantitative agreement with experiments was obtained. However,
interfaces are absent in the Gibbs ensemble~\cite{panagiotopoulos1987a}, so
these simulations do not enable investigations close to the critical
point, nor investigations of the interface in the two--phase region.

The general problem in the simulation of asymmetric binary mixtures (such
as the AO model) is that, by displacing or inserting a large particle,
overlap with a number of small particles will likely result. Such overlaps
are generally unfavorable and will be rejected in the majority of cases.
If no special steps are taken, one ends up displacing mainly small
particles while the large particles remain essentially frozen. For
asymmetric binary mixtures in the canonical ensemble a number of
specialized algorithms have been developed that circumvent this
problem~\cite{biben1996a,buhot1998a}. Unfortunately, it is difficult to
obtain the surface tension accurately in the canonical ensemble because it
must be derived from the rather small anisotropy of the pressure tensor in
that case: long wavelength interfacial fluctuations are hard to
equilibrate in the canonical ensemble~\cite{varnik2000a}.

In this letter we present a grand canonical Monte Carlo method that
enables direct simulations of asymmetric binary mixtures. We use the
method to study the fluid--fluid unmixing transition in the AO model. Our
method consists of collective Monte Carlo moves in conjunction with an
umbrella sampling technique that was recently developed by Virnau and
M\"uller~\cite{virnau2003a}. This way, we are able to calculate
the phase diagram of the AO model close to the critical point. At the same
time we can calculate the interfacial tension $\gamma$ in the two--phase
region using the method of Binder~\cite{binder1982a}. We present a
comparison of our data to recent DFT
results~\cite{brader2002a,schmidt2002a,schmidt2003a} and we show that the
AO model displays 3D Ising critical behavior.

The grand canonical ensemble requires that the temperature $T$, the volume
$V$ and the respective chemical potentials $\{ \mu_{\rm p}, \mu_{\rm c}
\}$ of the small (polymer) and large (colloid) particles are fixed. In a
standard grand canonical move one tries to insert or remove a particle
using Metropolis sampling~\cite{landau2000a}. This approach is not
efficient for asymmetric binary mixtures because the insertion of large
particles (colloids) is severely hampered by the presence of small
particles (polymers). The method that we present is aimed to circumvent
this problem. The main idea is to not transfer particles one--at--a--time,
but to swap clusters of small particles instead. This stimulates the
formation of voids, in which a large particle can be inserted without
producing overlap.

\begin{figure}
\includegraphics[width=60mm]{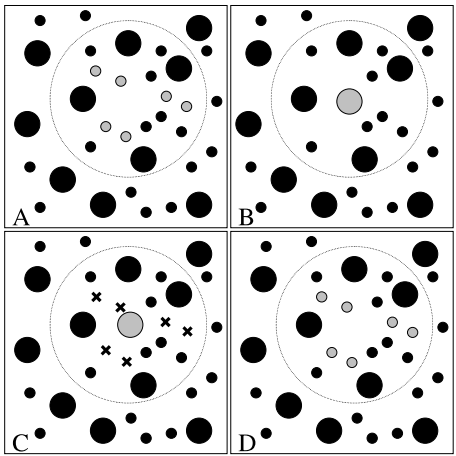}
\caption{~\label{figmove}Schematic picture of the grand canonical Monte
Carlo moves used in our simulations. Particles to be inserted or removed
are shaded grey. See text for details.}
\end{figure}

Figs.~\ref{figmove}A and \ref{figmove}B demonstrate the insertion
of an additional large particle into an asymmetric binary mixture
currently containing $N_{\rm c}$ large particles. First, a point
is selected randomly in the mixture and a sphere with radius
$\delta$ and volume $V_\delta=4\pi\delta^3/3$ is drawn around it (see
Fig.~\ref{figmove}A). Let $n_{\rm p}$ denote the number of small particles
inside the sphere: a particle is inside the sphere when the coordinates
of its center are inside the sphere. Next, we choose a uniform random
integer $n_{\rm r}$ from the interval $0 \leq n_{\rm r} < m$, with $m$
an integer that will be specified later. If $n_{\rm r} > n_{\rm p}$
the move is rejected but if $n_{\rm r} \leq n_{\rm p}$, $n_{\rm r}$
small particles are randomly selected from the sphere out of the $n_{\rm
p}$ present (these particles are shaded grey in Fig.~\ref{figmove}A).
The $n_{\rm r}$ selected particles are then removed from the mixture
and a large particle is inserted at the center of the sphere (see
Fig.~\ref{figmove}B). The new configuration is accepted with probability:
\begin{equation}
  A_+ = \min \left[1,
             \frac{z_{\rm c} V}{N_{\rm c}+1}
             \frac{ (n_{\rm p})! }{ (n_{\rm p}-n_{\rm r})! }
             \frac{ {\rm e}^{-\beta \Delta E} }
                  { (z_{\rm p} V_\delta)^{n_{\rm r}} } 
             \right],
  \label{eq1}
\end{equation}
with $\Delta E$ the potential energy difference between the initial and
the final configuration and $\{z_{\rm c},z_{\rm p}\}$ the fugacity of
large and small particles, respectively. The fugacity $z$ is related to the
chemical potential via $z = \exp(\beta\mu)$ with $\beta = 1/(k_B T)$ and
$k_B$ the Boltzmann constant.

The reverse move is illustrated in Figs.~\ref{figmove}C and
\ref{figmove}D. First, a large particle is selected at random and a sphere
with radius $\delta$ is drawn around the center of this particle. Next, a
uniform random integer $n_r$ from the interval $0 \leq n_r < m$ is chosen
followed by the selection of $n_r$ random locations inside the sphere.
These random locations are marked as crosses in Fig.~\ref{figmove}C. The
selected large particle is now removed from the mixture and $n_{\rm r}$
small particles are placed on the locations selected before (the newly
inserted particles are shaded grey in Fig.~\ref{figmove}D). The new
configuration is accepted with probability:
\begin{equation}
  A_- = \min \left[1,
              \frac{N_{\rm c}}{z_{\rm c} V}
              \frac{ (n_{\rm p})! (z_{\rm p} V_\delta)^{n_{\rm r}} }
              { (n_{\rm p}+n_{\rm r})! }
          {\rm e}^{-\beta \Delta E} \right],
  \label{eq2}
\end{equation}
the notation being the same as in Eq.~(\ref{eq1}).

It is straightforward to show that the acceptance probabilities $A_+$ and
$A_-$ enforce detailed balance, which ensures that the algorithm is not
statistically biased~\cite{newman1999a}. The algorithm is also ergodic
because single large particles have a finite probability of being inserted
anywhere in the system with one move. Similarly, a small particle can be
inserted anywhere via a combination of moves: for example, by the
insertion of a large particle followed by the removal of the same large
particle.

In order to apply the method to the AO model, the parameters $\delta$ and
$m$ still need to be specified. We use $\delta = R_{\rm c} + R_{\rm p}$
which is just big enough to contain one colloid in a sea of polymers. The
integer $m$ must be chosen large enough to allow for the formation of
voids. In the pure polymer phase, the polymer density equals $z_{\rm p}$
because the polymers behave like an ideal gas. The insertion sphere will
then contain $z_{\rm p} V_\delta$ polymers on average and thus we choose
$m$ slightly above this value. With this choice of $\delta$, the insertion
of a colloid can only succeed if all polymers are removed from the
insertion sphere $V_\delta$. This will in general happen a fraction $1/m$
of the time. To boost the acceptance rate, we choose to remove {\it all}
polymers from $V_\delta$ when we attempt to insert a colloid, provided
their number does not exceed $m$: moves that attempt to remove more than
$m$ polymers are rejected. To maintain detailed balance, the acceptance
probabilities $A_+$ and $A_-$ must be multiplied by $1/m$ and $m$,
respectively.

The phase diagram of the AO model can be expressed in terms of the reduced
polymer packing fraction $\eta_{\rm p}^{\rm r} \equiv (4\pi/3) z_{\rm p}
R_{\rm p}^3$ as a function of the colloid packing fraction $\eta_{\rm c}
\equiv (4\pi/3) R_{\rm c}^3 N_{\rm c}/V$. This is analogous to the
temperature--density phase diagram for simple fluids. In case of the AO
model, $\eta_{\rm p}^{\rm r}$ plays the role of inverse temperature and
$\eta_{\rm c}$ that of order parameter. In order to determine the
coexistence curve of the fluid--fluid unmixing transition, we calculate
for a given value of $\eta_{\rm p}^{\rm r}$ the probability distribution
$P(\eta_{\rm c})$.  This is the probability of observing a mixture with
colloid packing fraction $\eta_{\rm c}$. If phase separation occurs
$P(\eta_{\rm c})$ is bimodal: the peak at low $\eta_{\rm c}$ corresponds
to the colloid vapor phase, the peak at high $\eta_{\rm c}$ to the colloid
liquid phase, and the region in between is the phase--separated regime. To
ensure phase coexistence, the colloid fugacity $z_{\rm c}$ is tuned such
that the area under both peaks is equal~\cite{landau2000a}.

A crucial point in our simulation is the use of a new biased sampling
technique called successive umbrella sampling. This technique was recently
developed by Virnau and M\"uller~\cite{virnau2003a}. Combination of our
grand canonical Monte Carlo move and successive umbrella sampling enables
us to sample $P(\eta_{\rm c})$ also in regions where, due to the
free energy barrier separating both phases, $P(\eta_{\rm c})$ may be
very low.

\begin{figure}
\includegraphics[width=65mm]{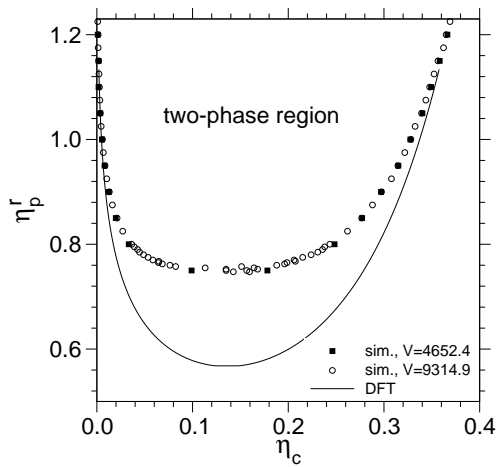}
\vspace{2mm}
\caption{~\label{figphase}Phase diagram of the AO model with $q=0.8$. The
points are the binodals as obtained from the simulation at the indicated
volumes of the simulation box. The solid line is the binodal from 
DFT~\cite{schmidt2002a,brader2002a}.}
\end{figure}

In the following we consider an AO mixture with a polymer to colloid size
ratio $q\equiv R_{\rm p}/R_{\rm c} = 0.8$. Note that all lengths are given
in units of $R_{\rm c} \equiv 1$. From previous studies we expect
fluid--fluid phase separation for $q=0.8$ in a wide range of $\eta_{\rm
p}^{\rm r}$. The simulations were performed in an elongated box with
aspect ratio $1/2$ and periodic boundary conditions. We have performed
simulations for two different volumes of the simulation box: $V_1=4652.4$
and $V_2=9314.9$. When the colloid packing fraction reaches $0.45$, the
smaller volume contains $500$ colloids and the larger volume $1000$
colloids.

Fig.~\ref{figphase} shows the phase diagram for the two system
sizes $V_1$ and $V_2$.  Comparison of the two data sets shows that
finite--size effects are relatively small, even close to the critical
point. Also included in Fig.~\ref{figphase} is the phase diagram obtained
from recent DFT~\cite{schmidt2002a,brader2002a}. We observe that DFT
underestimates the location of the critical point by about 30\%. More
importantly, DFT yields the typical mean--field parabolic shape of the
binodal (critical exponent $\beta=1/2$) while the simulation yields the
expected flatter binodal ($\beta \approx 0.325$; Ising model universality
class~\cite{luiten2001a}). We will demonstrate below that the simulation
indeed displays Ising critical behavior.

\begin{figure} \includegraphics[width=65mm]{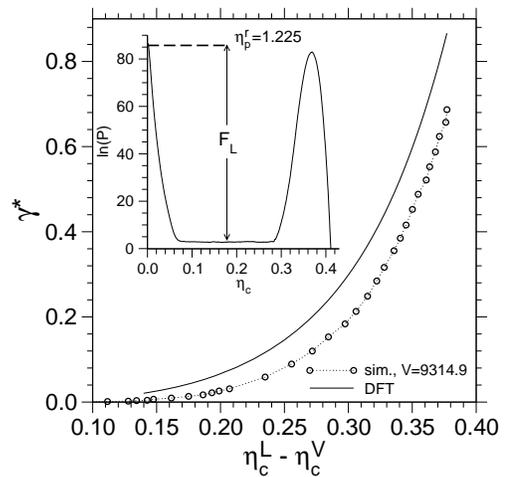}
\caption{~\label{figgamma}Reduced interfacial tension $\gamma^*
\equiv 4 R_{\rm c}^2 \gamma$ as a function of the difference between the
colloid packing fractions in the coexisting liquid (L) and vapor (V)
phases. The open circles represent our simulation data. The solid line
is a DFT result~\cite{schmidt2003a}. The inset shows the logarithm of
the probability distribution $P(\eta_{\rm c})$ for $\eta_{\rm p}^{\rm
r}=1.225$ with $F_{\rm L}$ defined in the text.} \end{figure}

The probability distribution $P(\eta_{\rm c})$ can also be used to extract
the interfacial tension $\gamma$ between the coexisting colloid vapor and
colloid liquid phases. To this end one uses a formula that was first
derived by Binder~\cite{binder1982a}:
\begin{equation}
  \gamma \equiv \lim_{L \to \infty} \frac{F_{\rm L}}{2 L^2} = 
                \lim_{L \to \infty} \frac{1}{2 L^2} 
                {\rm ln} \left[ 
                \frac{P^{\rm max}(\eta_{\rm c})}{P^{\rm min}(\eta_{\rm c})}
                \right],
\end{equation}
with $P^{\rm max}(\eta_{\rm c})$ and $P^{\rm min}(\eta_{\rm c})$ the value
of $P(\eta_{\rm c})$ at its maxima and its minimum, respectively, and $L$
the length of the simulation box parallel to the interface. An example
distribution is shown in the inset of Fig.~\ref{figgamma} for $\eta_{\rm
p}^{\rm r}=1.225$. Note that the presence of a flat region between the two
peaks is important for an accurate estimate of $\gamma$. This is enforced
in the simulation by using an elongated box.

Fig.~\ref{figgamma} shows the reduced interfacial tension $\gamma^* \equiv
4 R_{\rm c}^2 \gamma$ as a function of the difference between the colloid
packing fractions in the coexisting phases, as obtained from our
simulation together with a recent DFT result. Again, the DFT result
deviates from the simulation by about 30\%. This demonstrates that the
rather perfect agreement of DFT with experimental data as claimed in
Ref.~\cite{brader2002a} is coincidental. According to our simulation data,
the values of $\gamma^*$ for the AO model underestimate experimental data
significantly which shows that more sophisticated models are required to
describe colloid--polymer mixtures on a quantitative level.

\begin{figure}
\includegraphics[width=65mm]{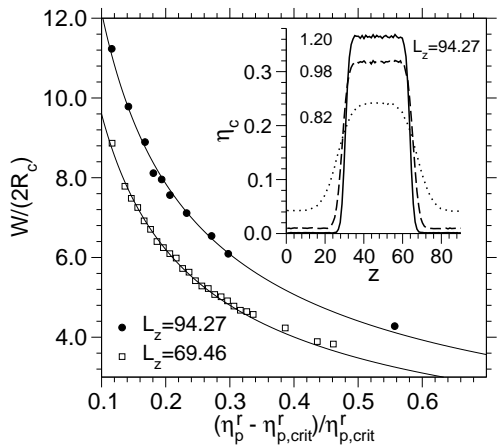}
\caption{~\label{fig3}The width of the colloid interface $W/(2 R_{\rm c})$
as a function of the relative distance from the critical reduced polymer
packing fraction for two different system sizes as indicated. The solid 
lines are fits to the 3D Ising power law described in the text. The inset 
shows three actual colloid density profiles along the direction 
perpendicular to the interfaces for $\eta_{\rm p}^{\rm r}= 
\{0.82,0.98,1.20\}$.}
\end{figure}

We have carried out additional simulations to calculate colloid density
profiles in the two--phase region. These simulations were performed using
$1156$ and $2889$ colloidal particles at a colloid packing fraction of
$0.13$. Periodic boundary conditions were again used and the aspect ratio
of both boxes was $1/3$. This corresponds to a box length of $L_{\rm
z}=69.46$ and $L_{\rm z}=94.27$ for the smaller and larger system,
respectively (here $L_{\rm z}$ is the box length perpendicular to the
interface). Three density profiles for the larger system are shown in the
inset of Fig.~\ref{fig3}. We have estimated the (10\%--90\%)-interfacial
width $W$ from these profiles by fitting the profiles to a hyperbolic
tangent. Since the interfacial width is expected to diverge with the
same exponent as the bulk correlation length near the critical point,
one expects $W \propto \left( \eta_{\rm p}^{\rm r}-\eta_{\rm p,crit}^{\rm
r} \right)^{-\nu}$, with $\nu=0.63$ corresponding to 3D Ising critical
behavior~\cite{luiten2001a}.  Fig.~\ref{fig3} shows that the data
for both system sizes is consistent with this power law. We can also
infer from Fig.~\ref{fig3} that $W$ depends strongly on $L_z$, even
far away from the critical point. This is most likely due to capillary
waves~\cite{binder2000} and care has to be taken when comparing $W$
from this simulation to interfacial widths obtained in experiments or
analytical theories.

In summary, we have presented a grand canonical Monte Carlo method which
is well suited to simulate asymmetric binary mixtures. It is
particularly powerful when combined with a re--weighting scheme: both the
phase diagram and the surface tension can be obtained accurately in that
case. We have used the method to determine the coexistence line of the
fluid--fluid transition in the AO model with high accuracy. We have
also presented new analysis of the interface between coexisting phases in
the AO model, in particular estimates of the interfacial tension.

We are grateful to the Deutsche Forschungsgemeinschaft for support
(TR6/A5) and to K. Binder, M. M\"{u}ller, M. Schmidt and P. Virnau for
stimulating discussion.

\bibliographystyle{prsty}

\end{document}